\documentclass[12pt]{article}
\title{Perturbative-nonperturbative interference  in
the static QCD interaction  at small distances}
 \author{Yu.A.Simonov\\Institute of Theoretical and Experimental
 Physics} \date{} \newcommand{\be}{\begin{equation}}
\newcommand{\ee}{\end{equation}} \def\la{\mathrel{\mathpalette\fun
<}} 
\def\fun#1#2{\lower3.6pt\vbox{\baselineskip0pt\lineskip.9pt
\ialign{$\mathsurround=0pt#1\hfil
##\hfil$\crcr#2\crcr\sim\crcr}}}

\begin{document}
\maketitle
\begin{abstract}

Short distance static quark--antiquark interaction is studied
systematically using the background perturbation theory with
nonperturbative background described by field correlators. A
universal  linear term $\frac{6N_c\alpha_s\sigma r}{2\pi}$ is
observed at small distance $r$ due to the  interference between
perturbative and nonperturbative contributions.
Possible modifications of this term due to additional
subleading terms are discussed and implications for systematic
corrections to OPE are formulated.
 \end{abstract}

\section{Introduction}

It is more than 20 years ago that the  power correction has been
computed in OPE [1] laying ground for numerous later
applications in QCD. Since then OPE  is the basic
formalism for study of short--distance phenomena, such as
DIS, $e^+e^-$ annihilation and, with some modifications, heavy quark
systems.

 Interaction of static charges at small distances has drawn
a lot of attention recently [2-4]. The theoretical reason is that
the appearance of linear terms in the static potential $V(r) =$ const
$r$, where $r$ is  the distance between charges, implies violation of
OPE, since const $\sim $ (mass)$^2$ and this dimension is not
available in terms of field operators.  There are however some
analytic [5,6] and numerical arguments [7,8] for the possible
existence of such terms $O(m^2/Q^2)$ in asymptotic expansion at large
$Q$.

Of special importance is the sign of the mass squared term. It was
argued recently in [3] that the small distance region may produce
tachyonic mass correction and this correction was studied
selfconsistently in different QCD processes. In particular the
correct (positive) sign of linear potential at small distances comes
from tachyonic gluon mass, while positive mass squared  term produces
negative slope of linear potential.

On a more phenomenological side the presence of linear term at small
distances, $r<T_g$, where $T_g$ is the gluonic correlation length
[9,10], is required by at least two sets of data.

First, the detailed lattice data [11] do not  support
much weaker  quadratic behaviour of  $V(r)\sim$ const $r^2$,
following from OPE and field correlator method [9,10], and instead
prefer the same  linear form $V(r) =\sigma r$ at all distances (in
addition  to perturbative $-\frac{ C_2\alpha_s}{r}$ term). Second,
the small--distance linear term is necessary for  the description of
the fine  structure  splittings in heavy quarkonia, since the
spin--orbit Thomas term $V_t=-\frac{1}{2m^2r}\frac{dV}{dr}$ is
sensitive to the small $r$ region and additional linear contribution
at $r<T_g$ is needed to fit the experimental splittings [12].
Moreover  the lattice calculations [13] display $1/r$ behaviour  for
the spin--orbit potential $V'_1$ in all measured region up to $r=0.1
fm$.

In what follows we display the basic dynamics which produces
tachyonic gluon mass and estimate its magnitude.

 \section{Background
perturbative theory}

 In this letter we report the first application of the
systematic background perturbation theory [14] to the problem in
question.  One starts with the decomposition of the full gluon vector
potential $A_\mu$ into nonperturbative  (NP) background $B_\mu$ and
perturbative field $a_\mu$,
\be A_\mu=B_\mu+a_\mu, \ee
 and the
'tHooft identity for the  partition function
\be
Z=\int DA_\mu
e^{-S(A)}= \frac{1}{N} \int DB_\mu\eta(B) \int Da_\mu e^{-S(B+a)}
\ee
where $\eta(B)$  is the  weight for
nonperturbative fields, defining the  vacuum averages, e.g.
\be
<F^B_{\mu\nu}(x) \Phi^B{(x,y)} F^B_{\lambda\sigma}(y)>_B=
\frac{\hat 1}{N_c} (\delta_{\mu\lambda}\delta_{\nu\sigma}-
\delta_{\mu\sigma}\delta_{\nu\lambda})D(x-y)+\Delta_1
\ee
where $F^B_{\mu\nu},\Phi^B$ are field strength and parallel
transporter made of $B_\mu$ only; $\Delta_1$ is the full derivative
term  [9] not contributing to string tension $\sigma$, which  is
\be
\sigma  =\frac{1}{2N_c} \int d^2 x D(x) +O(<FFFF>)
\ee

The background perturbation theory is an expansion of the last
integral in (2) in powers of $ga_\mu$ and averaging over $B_\mu$ with
the  weight  $\eta(B_\mu)$, as shown in (3). Referring the reader to
[14] for explicit formalism and  renormalization, we concentrate
below on the static interquark interaction at small $r$. To this end
we  consider the Wilson loop of size $r\times T$, where $T$ is large,
$T\to \infty$,  and define
\be
<W>_{B,a}= <P exp ig\int_C(B_\mu+a_\mu) dz_\mu>_{B,a}\equiv
exp\{-V(r)T\}
\ee
Expanding  (5)  in powers of $g a_\mu$, one obtains
\be
<W>=W_0+W_2+...;~~V=V_0(r)+V_2(r)+V_4(r)+,
\ee
where $V_n(r)$  corresponds to  $(g a_\mu)^n$ and can be expressed
through $D,\Delta_1$ and higher correlators [10,14] and its behaviour
at small $r$ is [9]
\be V_0(r)=C_0 r^2+C'_0r^4+...,~~ r\la T_g, V_0(r) =\sigma r,~~ r\gg
T_g
\ee
where coefficients are integrals of field
correlators over Euclidean time,
$$
C_0=O(<FF>) =\int^\infty_0D(\nu) d\nu  +O(\Delta_1),  C'_0=O(<F^4>).
$$
It is this small $r$ behaviour (7) which causes phenomenological
 problems mentioned above.

Coming now to $V_2(r)$, describing one exchange  of perturbative
gluon in the background, one finds from the quadratic in $a_\mu$ term
in $S(B+a)$ in the background Feynman gauge the gluon Green's
function
\be
G_{\mu\nu} =-(D^2_\lambda \delta_{\mu\nu}+ 2ig F^B_{\mu\nu})^{-1},
~~D_\lambda^{ca}=\partial_\lambda \delta_{ca}+g f^{cba}
B^b_\lambda
\ee
The  term $W_2$ can be  written through $G_{\mu\nu}$ as
\be
W_2=g^2\int^T_0dx_4\int^T_0 dy_4<P
exp(ig\int_CB_\mu dz_\mu) G_{44}(x,y)>_B
\ee
where for $G_{44}$ one can use  the  Feynman--Schwinger
representation (FSR) [10]. The simplest form of FSR obtains when one
can neglect or expand in powers of gluon spin interaction
(paramagnetic term $2ig F_{\mu\nu}^B$ in (8)).

Doing this expansion, $G_{\mu\nu}$ can be written as
\be
G=-D^{-2}+D^{-2} 2ig F^B D^{-2}- D^{-2} 2ig F^BD^{-2} 2ig F^BD^{-2}
\ee
the first   term on the r.h.s. of (10) corresponds to the spinless
gluon exchange in the background $B_{\mu}$, which can be written
using FSR [10] as
\be
G_{44}(x,y)=\int^\infty_0 ds(Dz)_{xy}e^{-K} Pexp ig\int_{C(z)} B_\mu
dz_\mu,~~K=\frac{1}{4}\int^s_0\dot z^2 d\tau
\ee

Here $B_\mu$ is in the adjoint representation of SU($N_c$). At large
$N_c$ one can use the 'tHooft rule to replace the gluon adjoint
trajectory $C(z)$ by the double fundamental trajectory which forms
together with the  original rectangular contour $C$ in (9) two closed
Wilson loops (see [14] for details and discussion).

The  average over $B_\mu$ in (9) then reduces (again in the large
$N_c$ limit) to the product of two averaged Wilson loops, namely

\be
W_2= \int^\infty_0 ds (Dz)_{xy}e^{-K}dx_4 dy_4 <W(C_1)><W(C_2)>,
~~K=\frac{1}{4} \int^s_0\dot z^2d\tau,
 \ee
  where $C_1$  and $C_2$ are
two contours obtained from the rectangular Wilson loop when two
points on it, $x$ and $y$, are connected by a double line of gluon
trajectory. It is convenient to choose the surfaces $S_i$ inside
$C_1$ and $C_2$ as consisting  of two adjacent pieces
 $S_i(C_i)=S_i(plane)+S(\Delta),$ one lying on the plane of original
Wilson loop and another piece $\Delta$, perpendicular to the plane
and bounded by the trajectory. Using now the nonabelian Stokes theorem
[9] and cluster expansion for the  average $<W(C_i)>$, one obtains
that bilocal correlator of  fields with points on the  two pieces
vanishes since $<E_iE_k>\sim \delta_{ik}$, and $<E_i
B_k>=0(\Delta_1)$ and vanishes by symmetry arguments.
Trilocal and higher correlators for dimensional resons bring with
them higher powers of distance $r$ and can be neglected.

Hence one has
\be
<W(C_1)><W(C_2)>
=<W_\Delta>^2 W_0
\ee
where
\be
W_\Delta = exp (-\frac{1}{2N_c}\int_{\Delta} D(x-y)
d\sigma_{\mu\nu} (x)  d\sigma_{\mu\nu}(y))
\ee

Two different regimes are possible for (14). In the small distance
region, $r\la T_g$, the sizes  of the surface $\Delta$ are of the
order of $r$ and one can replace $D(x-y)\to D(0)$ in (14),
and for dimensional reasons
the only possible contribution is
\be
W_\Delta \cong 1+O(D(0)r^4)
\ee
Thus one obtains a correction $O(r^4)$ to the perturbative potential
$1/r$, and hence no linear term.

In the large distance region, $r\gg T_g$, one obtains the area law
for $W_\Delta$,
\be
W_\Delta\cong exp (-\sigma S_\Delta)
\ee
Insertion of (16) in (12) yields  a massive propagator of
a spinless hybrid with mass $m=m_\Delta$ at large $r$,
 which corresponds to the
first excitation of the open string with fixed ends.

Summaring one can rewrite $W_2$  as
\be
W_2=W_0\int dx_4\int dy_4 <G(x,y)>
\ee
where $G(x,y)>$
is the Green's function of the  spinless hybrid. One can satisfy the
properties (15), (16)  representing $<G>$ as the propagator of a
particle with variable mass $m_0(p)$,
\be
<G(p)>=\frac{1}{p^2+m^2_0(p)}
\ee
where $m^2_0(p)\la O(\frac{1}{p^2}),p\to\infty$, and $m^2_0(p\to 0) =
m^2_\Delta.$

Thus at small distances (large $p$) $<G(p)>$ describes the usual
massless  gluon exchange, whereas at large distances it describes the
propagation of the spinless hybrid.

Hence at small $r\la T_g$ the background field $B_\mu$ in $D^{-2}$
is not operative and one can replace $D^{-2}$ by the free gluon
propagator $\partial^{-2}$. The contribution of the second term to
$W_2, dx_\mu D^{-2} F_{\mu\nu}^B D^{-2} dy_\nu$, obtains
when $x$ and $y$ are on adjacent sides of the Wilson loop and
therefore does not affect $V(r)$. In what follows we concentrate on
the third term in (10), $W_2^{(3)}$,

$$
W_2^{(3)} ={4g^2} \int^T_0 dx_4 \int^T_0 dy_4
<G(x,u)> d^4u< g^2 F^B_{4i} (u) F_{i4}^B(v)>
$$
\be
\times <G(u,v)> d^4v <G(v,y)>
\ee

One can rewrite (19) as
\be
W_2^{(3)} = 2g^2 \int^T_0 dx_4 \int^T_0 dy_4 \int \frac{d^4
ke^{ik(x-y)}}{(k^2+m_0^2(k))^2(2\pi)^4} \lambda^2(k),
\ee
where we have defined,
having in mind (4)
\be
\mu^2(k) =3\int \frac{D(z) e^{-ikz} d^4z}{4\pi^2 z^2};
 \mu^2(0) = \frac{3\sigma N_c}{2\pi},
\ee
Doing integrals over $dx_4  dy_4$, one gets
 \be
 W^{(3)}_2 = \frac{T}{\pi^2}\int \frac{d^3 ke^{i\vec
 k\vec r}\alpha_s(k)\lambda^2(k) }{(\vec k^2+ m_0^2(k))^2}= -\Delta
 V_2(r)T
 \ee
 To estimate the integral (22) one can take $\alpha_s(k)\lambda^2(k)$
 out of integral at some effective point $k_0$ and calculate the rest
 in a simple way, assuming $m_0$ to be constant, and expanding result
 at small $r$.

 In this way one obtains
 \be
 \Delta V_2(r)
 =\frac{\alpha_s(k_0)\lambda^2(k_0)}{m_0r}\frac{\partial}{\partial
 m_0} e^{-m_0r}= \alpha_s(k_0) \mu^2(k_0) (-\frac{1}{m_0}+r+O(r^2) )
 \ee
 Analysis of (21) tells that $\lambda^2(k)$ is a rather weak function
 of the argument, and to get an idea of the magnitude of $\Delta
 V_2(r)$, one can approximate $\lambda^2(k_0)\cong \lambda^2(0)
 =\frac{3\sigma N_c}{2\pi}$ yielding  for $\Delta V_2(r)$,

 \be
 \Delta V_2(r)\sim
 \frac{3N_c \alpha_s\sigma r}{2\pi}, ~~ r\la T_g.
 \ee
 Note, that had we renormalized $\alpha_s$
 in (23), (24) in the standard way
 we would
 meet the IR divergence of the  running $\alpha_s(k)$, since
 the corresponding momentum $k_0$  is in the IR regime
 for the constant term $(-\frac{1}{m_0})$ in (23).
 However, the NP background formalism predicts IR modification  of
 $\alpha_s$ (see [14,15] for details and discussion),  the
 so-called freezing $\alpha_s$ behaviour, which from heavy quarkonia
 fitting was found in [16] to yield maximal $\alpha_s$ at small $k$,
 \be
  \alpha_s(max)=0.5
   \ee
    For the linear term in (23) the situation
    is different\footnote{The author is grateful to V.I.Zakharov for
    the discussion of this point}
    and the effective value of $k_0$ is of the order of $1/r$. Hence
    it is more appropriate to present (23) in the form
    \be
    \Delta V_2(r)=\alpha_s(1/r)\mu^2(1/r) r
    \ee

     One
  may wonder  whether behaviour (24) holds also at large $r$, thus
  increasing linear potential $V_0(r)$. However the form (24) was
  obtained at small $r\la T_g$, while at large $r$ next terms of
  expansion in (10)  are contributing
   and the whole series (10) should be summed up explicitly.

  One can perform the summation, replacing  first for simplicity
   $D\to \partial$ in (8), (small $r$ or relatively large
  $q$ are considered) and assuming that only bilocal correlators of
  $F_{\mu\nu}$ are nonzero. One obtains
  $$
  G=-(\partial^2+2ig F)^{-1}\to G(k) =\frac{1}{k^2-\mu^2(k)},$$
  where $\mu^2(k)>0$. Thus $G(k)$ acquires a pole  at
  real value $\mu^2(k)$ in Euclidean space-time, signalling
  appearance of a tachyon. From physical point of view this result is
  a consequence of paramagnetic attractive interaction of gluon spin
  with NP background, yielding negative correction to the gluon
  selfenergy. A similar  term occurs for a quark due
  to its spin interaction with background [17].

    Recently a negative (tachyonic) mass shift was observed due to
    the gluon interaction with the stochastic background in [18].

  It is meaningful, that the same paramagnetic term $F^B_{\mu\nu}$ in
  $G_{\mu\nu}$ yields  negative contribution to the charge
  renormalization  (asymptotic freedom) [19].
   In fact negative paramagnetic effective action $S_{eff}^{(para)}$
   [19] gives rise to the negative (tachyonic) mass since both are
   connected, $-\mu^2\delta_{\mu\nu}\sim \frac{\delta^2 S}{\delta
   a_\mu\delta a_\nu}$. Therefore one may expect that the phenomenon
   of tachyonic gluon mass is pertinent to  nonabelian theories.

    The
   existence of the tachyon reveals paramagnetic
  instability of the object (gluon or quark), if
  stabilizing mass is not created
  by some additional mechanism. In our case this mechanism is the
  creation of hybrid mass due to the same confining correlator $D(x)$
  when $D^2$ is used and not $\partial^2$ in
  (8). As was explained above, the hybrid mass is created at larger
  distances, $r\gg T_g$, so  that for  illustrative purposes
  (referring the hybrid mass to the gluon in question), one may write
  the total gluon propagator $G(k)$  as
  $$
   G(k)=\frac{1}{k^2+m^2(k)},
  $$
   where $m^2(k)= m^2_0(k)-\mu^2(k),$ and $m^2_0(k)$ is dominating
  at small $k$ (large distances) while $\mu^2(k)$ dominates at large
  $k$ (small distances). A simple example is provided by
  $m^2(k)=\mu^2\frac{\mu^2-k^2}{\mu^2+k^2},$ in which case one can
  calculate gluon  exchange potential $V(r)=-\int\frac{d^3\vec
  k}{(2\pi)^3}e^{i\vec k \vec r} G(k)$ explicitly to yield at small
  $r$:
   $$
   V(r)\sim-\frac{1}{r} + \frac{\mu^2}{2} r, ~~\Delta V\sim
  \frac{\mu^2}{2} r
   $$
    in agreement with the form (24).  At large
  distances one should calculate the exact gluon Green's function (8)
    in the Wilson loop $W^{(2)}$, which describes to the propagation
     of the hybrid state with two static quarks $Q\bar Q$ at the ends
    of the string.

  Such a state was considered both analytically [20] and on the
  lattice [21] yielding  excitation energy
  (which corresponds to the mass $m(0)$) around  1 GeV.

  Hence (24) is only a small distance approximation of the
  hybrid exchange potential, where the dominant paramagnetic
  contribution is kept in the effective gluon mass.

  \section{Other possible corrections to $V(r)$}

  In doing perturbative expansion in (2) one encounters other terms
  in $S(B+a)$ which potentially yield interference contributions of
  perturbative $a_\mu$ and NP $B_\mu$ fields. Of special importance
  is the term $L_1$,
  \be
  L_1=\int a_\nu D_\mu (B) F^B_{\mu\nu} d^4 x
  \ee
   Correction due to (27) in the gluon propagator was studied in
   Appendix 1 of [14] and  can be written in the form (20)
   where $\mu^2(k^2)\to J_1$ is now  expressed as
   \be
   J_1(k^2) =\int d^4 z e^{-ikz}<DF^B(z) DF^B(0)>
   \ee
   Now the integrand in (28) can be written as a sum of two terms
   [22],
   \be
   <D_\rho F^B_{\rho\nu}(z) D_\lambda F_{\lambda \mu}^B(u)>=
   \partial_\rho\partial_\lambda< F^B(z) F^B(u)>+0(<FFF>)
   \ee
   The first term on the r.h.s. of (29) yields $\delta J_1\sim k^2$
   and hence only a NP correction to the Coulomb term, while the
   second term, $<FFF>$, would give, using dimensional arguments,
   correction $\Delta V\sim <FFF> r^5$, negligible at small $r$.

   Another type of correction occurs from the interplay of multiple
   color Coulomb exchanges and one NP contribution, considered in
   [23]. This effect can be accounted for by the replacement
   \be
   D(x,t) \to D(x,t) exp (-\frac{N_c\alpha_s t}{2x})
   \ee

   Here $t$ is the Euclidean time and the exponent in (30) accounts
   for the difference of potential in singlet and octet channels.
   Insertion of (30) in (7) yields an additional suppression of the
   $r^2$ dependence. This result coincides with the correction
   obtained in [24] in a different way.

   Finally we consider in this section the correction due to the
   freezing behaviour of the coupling $\alpha_s$ [14,15],
   \be
   \alpha^f_s(r)
   =\frac{4\pi}{b_0ln\frac{1+r^2m^2_B}{\Lambda^2r^2}}\cong
   \alpha_s^{(0)}(r) -\frac{\alpha_s^{(0)} m_B^2
   r^2}{ln\frac{1}{\Lambda^2r^2}}
   \ee

   As was noticed by F.J.Yndurain [2], expansion of the freezing
   Coulomb potential yields a linear term
   \be
   \Delta V_c(r)=+\frac{C_2\alpha_s^{(0)}
   m_B^2r}{ln\frac{1}{\Lambda^2 r^2}}+...
   \ee
   Here $m_B$  is the double hybrid mass, ($m_B=1.1 GeV$ from the fits
   to experiment [16]), and the  term  (32) is always much
   smaller than the Coulomb term because of  condition $ m^2_B r^2\ll 1$.
   One may consider $\Delta V_c$  in (32) as coming from the
   additional $1/p^2$ in $\alpha_s(p)$ , as was suggested in [8], but
   here  there  is the log term in the denominator,
   reminding that the pole is coming from the expansion of the
   freezing $\alpha_s(p)$.

   One should note that there is no double counting in adding (24)
   and (32), since (24) is obtained from the one-gluon exchange
   (OGE) process, while (32) is due to the one--loop corrections to
   OGE. However the region of validity of (32) is always smaller
   than (24).

   \section{Discussion  and conclusion}

   The analysis done   heretofore  concerns static  interquark
   potential and reveals that even at small distances NP background
   ensures some contributions which come from relatively  small
   intermediate
   distances, $l\la T_g$, and encoded in the  mass  squared term
    \be
    \mu^2=\frac{3\sigma N_c}{2\pi},\mu \approx 0.5
   GeV
   \ee
   which generates the  potential (24).

Applying the same NP background formalism to other processes of
interest, one would  get similar corrections of the order of
$\frac{\mu^2}{p^2}$, as e.g. in OPE for $e^+e^-$ annihilation.

One example of this kind is the calculation of the field correlator\\
$<F_{\mu\nu}(x) \Phi(x,y) F_{\lambda\sigma} (y)>$.
The leading contribution can be written using the gluon propagator
$G$ (8),
\be
<F\Phi F> = \partial_\mu\partial_\lambda G_{\nu\sigma} + perm +
G_{\mu\lambda} G_{\nu\sigma} +perm
\ee
where we have suppressed $\Phi$ and perm. denotes terms obtained by
the permutation of indices with the proper change of  sign.
Insertion of expansion (10) into (34) yields in addition to the
standard perturbative term $O((x-y)^{-4})  $  a contribution
proportional to (33),
\be
<F(x) \Phi F(0)> \sim \frac{C_1}{x^4} +\frac{C_2\mu^2}{x^2}+...
\ee
where both $C_1$ and $C_2$ are positive computable numbers.

A recent lattice  study of a similar quantity [7] as  a function of
$UV$  cut-off $\Lambda$ reveals the possible presence of the
$O(\Lambda^2\sim x^{-2})$ term.

It is clear that appearance of $\mu^2$, which is an integral of
nonlocal entity $<F(x)F(0)>$ over a NP scale, $x\sim T_g$, violates
the original OPE of Wilson [25], proved in the pure perturbation
theory, and the extended OPE of Shifman, Vainshtein and Zakharov [26],
where NP contributions enter as matrix elements of  local operators.

This extended form of OPE   can be considered as a
 physically motivated assumption, and an  explicit
treatment done here  within the background perturbation theory (BPT)
reveals that some extra terms should be added to OPE, the first of
which, $\mu^2/p^2$, was discussed  in [3].

One might ask at this point, how rigorous and selfconsistent is BPT,
with nonperturbative background given by correlators. One should
stress here, that BPT is a consistent and systematic method, but not
a complete one, since no recepee  was suggested above   for
calculation of  NP correlators, and the NP configurations
are introduced by hand just as it is done in QCD sum rules [26].

However recently the situation has changed. In [27] equations have
been derived in the limit of large $N_c$ for vacuum correlators, from
which correlators can be computed one by one explicitly. In this way
the exponential form  of the lowest correlator, $<F(x)\Phi F(0)>$
was defined analytically [27] in agreement with lattice studies, and
analytic connection between $T_g$ and $\sigma$ was found,
yielding $T_g$   in a good agreement with lattice data [28].

The main conclusion from these studies is that NP configurations
appear as a selfconsistent solution of nonlinear equations which
violates spontaneously scale symmetry pertinent to these equations
and their perturbative solutions.

>From this point of view the NP background exploited here
can be identified with scale violating NP solutions in [27],
 and the BPT method is made complete.

On the phenomenological side the  account of the correction (24) and
(32) in the total potential
 $$ V(r)=V_0(r)-\frac{C_2\alpha_s^f(r)}{r}
  +\Delta V_2(r)
   $$
   may provide for $V(r)$ a simple "linear plus
  Coulomb" picture for all distances which is in better agreement
both with experiment [12,17] and with  lattice data [11, 13].

To conclude: in this study perturbative -- nonperturbative
interference was shown to provide additional OPE terms, absent in the
usual local OPE form. In addition there are purely NP contributions
[14] which are also outside of the standard lore, and will be
discussed elsewhere.

The author  is grateful  to V.A.Novikov and  V.I.Shevchenko for
fruitful discussions, and to V.I.Zakharov for discussions,
correspondence and very useful remarks.

The financial support of RFFI through the grants 97-02-16406 and
97-0217491  is gratefully acknowledged.

\end{document}